\documentclass[aps,
pra,superscriptaddress,
reprint,
amsmath,
amssymb
]{revtex4-1}
\usepackage[english]{babel}
\usepackage{amsmath,graphicx,color,bm}

\begin{document}
\title{Chiral topological phases in optical lattices without synthetic fields}

\author{Hoi-Yin Hui}
\affiliation{Department of Physics, Virginia Tech, Blacksburg, Virginia 24061, USA}
\author{Mengsu Chen}
\affiliation{Department of Physics, Virginia Tech, Blacksburg, Virginia 24061, USA}
\author{Sumanta Tewari}
\affiliation{Department of Physics and Astronomy, Clemson University, Clemson, South Carolina 29634, USA }
\author{V. W. Scarola}
\affiliation{Department of Physics, Virginia Tech, Blacksburg, Virginia 24061, USA}

\begin{abstract}
Synthetic fields applied to ultracold quantum gases can realize topological phases that transcend conventional Bose and Fermi-liquid paradigms. 
Raman laser beams in particular are under scrutiny as a route to create synthetic fields in neutral gases to mimic ordinary magnetic and electric fields acting on charged matter.  Yet external laser beams can impose heating and losses that make cooling into many-body topological phases challenging.  We propose that atomic or molecular dipoles placed in optical lattices can realize a topological phase without synthetic fields by placing them in certain frustrated lattices.  We use numerical modeling on a specific example to show that the interactions between dipolar fermions placed in a kagome optical lattice spontaneously break time reversal symmetry to lead to a topological Mott insulator, a chiral topological phase generated entirely by interactions.  We estimate realistic entropy and trapping parameters to argue that this intriguing phase of matter can be probed with quantum gases using a combination of recently implemented technologies. 
\end{abstract}

\maketitle

\section{Introduction}
Condensates of neutral atoms and molecules offer considerable opportunities for realizing quantum many-body states of matter\cite{Bloch2008}.  But conventional wisdom asserts that because they are not charged, these condensates must be manipulated by synthetic fields, instead of ordinary electromagnetic fields, to engineer quantum states beyond conventional Bose and Fermi liquids \cite{Lewenstein2007,Dalibard2011,Galitski2013,Goldman2009,Goldman2010,Goldman2016a}.   Methods to implement synthetic fields include external Raman lasers to yield effective magnetic fields or spin-orbit coupling.  

Strong synthetic fields can generate topological states in quantum degenerate gases.  A strong magnetic field leads to the integer quantum Hall effect by explicitly breaking time-reversal symmetry to allow chiral edge modes with suppressed back scattering that surround the otherwise insulating integer quantum Hall state \cite{Thouless1983}.   Similarly, strong spin-orbit coupling can lead to topological insulators with suppressed backscattering in edge modes \cite{Hasan2010,Qi2011}.  In both cases the synthetic field could, if realized, encode a topological invariant (an integer Chern number for the quantum Hall state or a $\mathcal{Z}_2$ invariant for the topological insulator) in the single-particle band structure that manifests in quantization of edge mode observables.  In spite of recent progress in realizing synthetic fields in ultracold atomic gases \cite{Lin2011,Cheuk2012,Wang2012b,Wu2016a, 
Lin2009,Lin2009b,Aidelsburger2011,Leblanc2012,Struck2013,Beeler2013,Aidelsburger2013,Miyake2013,Jotzu2014,Atala2014,Stuhl2015,Kennedy2015,Tai2017,Eckardt2017}, there has been concern that light-induced synthetic fields will ultimately suffer from heating and losses \cite{Stenger1999} (particularly in strongly-interacting regimes of interest \cite{Wei2013,Williams2013}) thus complicating the realization of topological states.

Topological phases, with quantized observables, can be realized even without synthetic fields.  For example, work in the context of chiral spin liquids \cite{Wen1989,Wen1991} pointed out that strong interactions can \emph{spontaneously break time reversal symmetry}, thus allowing a state with chiral edge modes and a non-zero Chern number even in the absence of an external magnetic field. A subsequent mean-field theory (MFT) work \cite{Raghu2008} on spinless fermions hopping in a honeycomb lattice (fermions with a linearly crossing band structure) suggested that interactions lead to a similar state: a topological Mott insulator phase (TMIP), with a quantum anomalous Hall (QAH) effect that arises solely from interaction effects.   The exciting possibility of finding materials with a QAH effect spurned more rigorous numerical studies that unfortunately  suggest that the TMIP in honeycomb lattice models is barely stable against competing charge density wave (CDW) order, if at all \cite{Capponi2015,Motruk2015,Kurita2016}.  

\begin{figure}
\begin{centering}
\includegraphics[width=0.85\columnwidth]{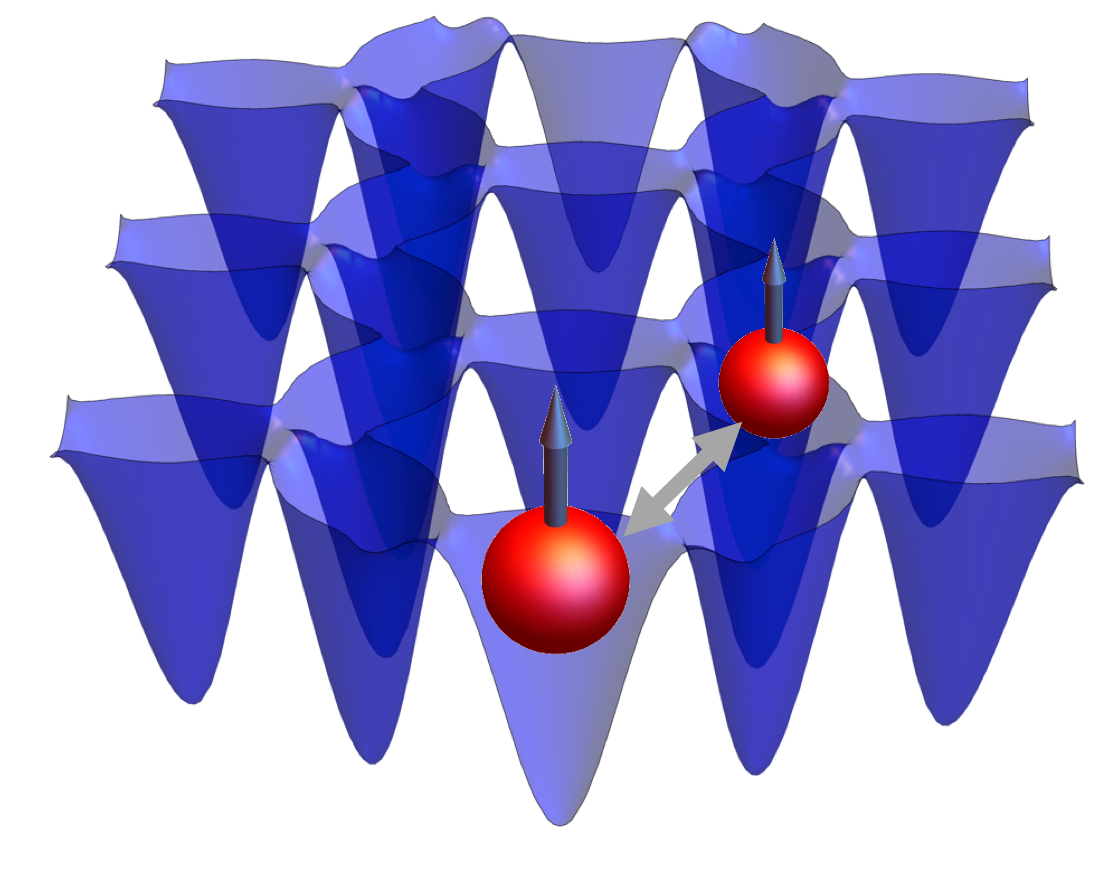}
\par\end{centering}
\caption{Plot of a kagome optical lattice potential \cite{Jo2012} (see Appendix~\ref{app:kagomeFormula} for the explicit formula) as a function of position in the $x-y$ plane.  The two particles represent schematics of dipoles separated in the plane by $\vert {\bm r}-{\bm r}'\vert$  with moments oriented perpendicular to the plane to ensure mutual repulsion. }
\label{fig:opticallattice}
\end{figure}

Recent works show that interacting models favor a TMIP in certain frustrated lattices with a quadratic band crossing point (QBCP) \cite{Sun2009,Zhang2009,Wen2010,Chen2017} rather than a linear crossing point as in the honeycomb lattice.  QBCPs arise in several frustrated two-dimensional lattices, including the kagome, diamond, Lieb, and decorated honeycomb lattices. MFT modeling of short-range interacting  fermions on the kagome lattice shows evidence for a TMIP \cite{Wen2010}.  Density matrix renormalization group studies on a kagome lattice with an inverted band structure (positive tunneling energy between sites rather than the usual negative tunneling energy found in optical lattices) also show strong evidence for a TMIP for short-range interactions \cite{Zhu2016}.  While these theory works are promising, it is still unclear how electronic matter can be coaxed into the TMIP since the electrons need to be fully polarized and the role of the long-range part of the electron-electron interaction remains an open issue.  

We propose that quantum degenerate gases of atomic or molecular dipoles \cite{Griesmaier2005,Stuhler2005,Chicireanu2006,Lu2012,Aikawa2012,Depaz2013,Aikawa2014, Ospelkaus2008a,Deiglmayr2008a,Pilch2009a,Voigt2009,Ni2010,Ospelkaus2010,Chotia2012,Wu2012,Heo2012,Yan2013} placed in a kagome optical lattice (Fig.~\ref{fig:opticallattice}) can be tuned to regimes that realize a topological Mott insulator with an observable QAH effect.  We combine complimentary methods [MFT and exact diagonalization (ED)] to 
show that the dipolar interaction supports a TMIP in a physically realistic tight-binding model of dipoles in a kagome optical lattice.  Our proposal avoids the need for synthetic fields and instead relies on technology recently implemented in experiments that have, separately, realized a kagome optical lattice \cite{Jo2012} and cooled dipolar gases to quantum degeneracy, e.g., $^{161}{\rm Dy}$ \cite{Lu2012}.  We estimate that the entropy required to reach the topological phase is $\sim 0.15 k_{\text{B}}$ per particle, potentially within reach of cooling capabilities with atomic gas microscopes \cite{Mazurenko2017}.  Our work therefore shows that a combination of recently implemented technologies with atomic and molecular condensates can be used to realize and observe a topological Mott insulator.

This paper is organized as follows. In Sec.~\ref{sec:model} we set up the problem and describe our ED and MFT methods. In Sec.~\ref{sec:results} we present the results of our analysis, demonstrating the emergence of the TMIP under suitable conditions. Finally in Sec.~\ref{sec:discussion} we summarize our findings with a discussion of prospects for experimentally realizing the TMIP with ultracold atoms in optical lattices. The appendices discuss the explicit formula of the optical lattice potential, the effects of finite spread of Wannier functions on the dipolar interaction, and the role of finite size effects in the calculation of current.

\section{Model and Methods}\label{sec:model}
We consider an optical lattice defined by three bichromatic laser beams intersecting at $120^\circ$ to define a kagome pattern\cite{Jo2012} (Fig.~\ref{fig:opticallattice}).  For a sufficiently deep optical lattice we may safely assume that all particles reside in the lowest three Bloch bands.  If the optical lattice is loaded with fermionic dipoles (with their dipolar moment aligned perpendicular to the plane), we may model the dipoles with the following tight-binding Hamiltonian:
\begin{equation}
H=-t\sum_{\left\langle \bm{r,r'}\right\rangle}\left(c_{\bm{r}}^{\dagger}c_{\bm{r'}}^{\vphantom{\dagger}}+{\rm H.c.}\right)+\frac{V_1}{2}\sum_{\bm{r} \neq \bm{r}'}\frac{n_{\bm{r}}n_{\bm{r'}}}{|\bm{r}-\bm{r'}|^3},
\label{eq_H}
\end{equation}
where $c_{\bm{r}}^{\vphantom{\dagger}}$ ($c_{\bm{r}}^{\dagger}$) annihilates (creates) a spinless fermion at the site $\bm{r}$ and $n_{\bm{r}}=c_{\bm{r}}^{\dagger}c_{\bm{r}}^{\vphantom{\dagger}}$.
The first term is the single-particle tunneling between neighboring sites.  In the following we work in units with $t=k_{\text{B}}=1$.  We also set the nearest-neighbor lattice spacing to unity.

The last term in Eq.~\ref{eq_H} approximates the dipolar interaction.  The prefactor $V_1$ is the interaction energy between nearest neighbors.  The interaction is written in the limit of infinitely narrow Wannier functions. Corrections to this interaction derived from the finite spatial extent of the Wannier functions are discussed in Appendix~\ref{app:dipolar}.  We find that realistic corrections to the interaction term do not significantly impact our findings. 

To numerically study Eq.~\ref{eq_H}, we truncate the interaction when the interaction strength becomes weak so that the truncation does not significantly  impact our results.  In our mean-field results carried out in the thermodynamic limit, the interaction includes all pairs of sites with 
$|\bm{r}-\bm{r'}|<5$.  In our finite size studies (where we compare MFT and ED) the interaction includes pairs only up to $|\bm{r}-\bm{r'}|<2$ to avoid finite size effects. 

\begin{figure}
\begin{centering}
\includegraphics[width=1\columnwidth]{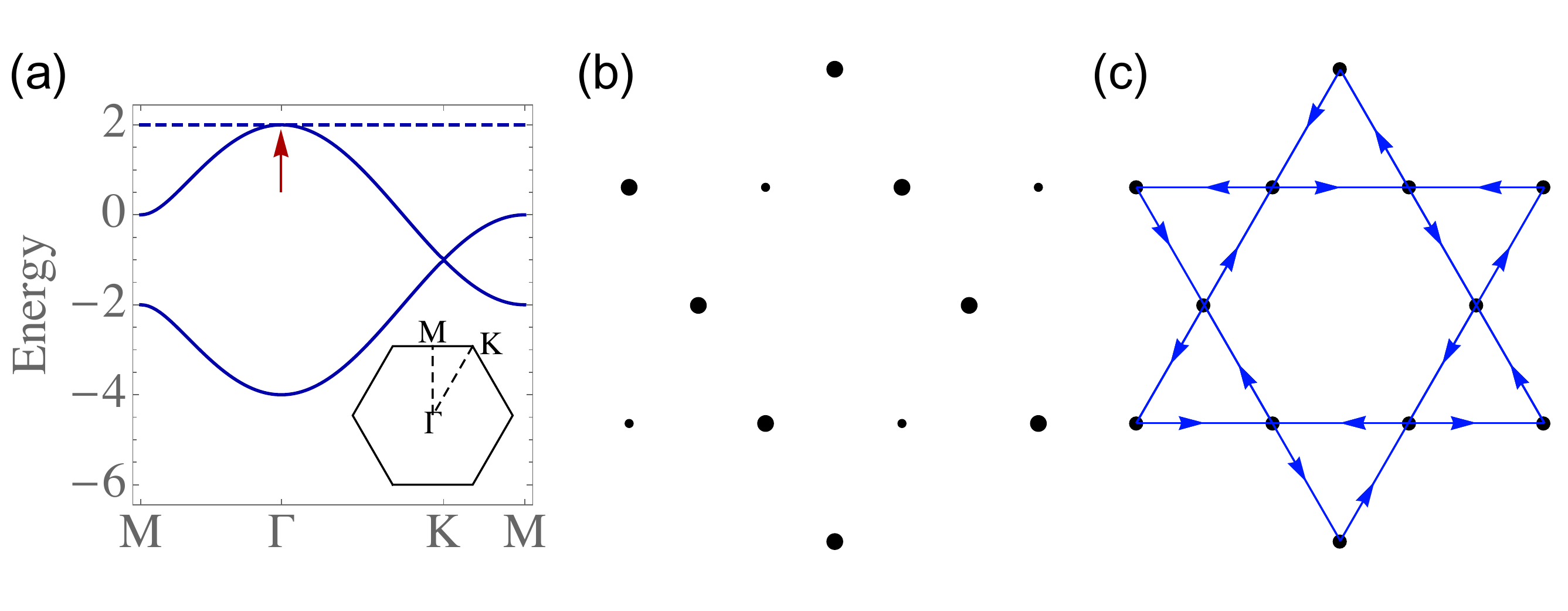}
\par\end{centering}
\caption{(a) Single-particle energies as a function of wavevector on a kagome lattice with only nearest-neighbor tunneling.
At a density of $2/3$, the bands marked with solid (dashed) lines are filled (empty),
and the red arrow shows where the Fermi surface touches the empty band at the quadratic band crossing point. The inset shows the definition of various high-symmetry points in the first Brillouin zone.
(b) The charge density wave pattern obtained from Eq.~\ref{eq_H} with $V_1=2$. The sizes of the dots are proportional to the average occupation number.
(c) The chiral current pattern in a topological Mott insulator phase with a quantum anomalous Hall effect obtained from Eq.~\ref{eq_H} with $V_1=1.8$.  Exact diagonalization and mean-field theory obtained the same patterns found in both (b) and (c). 
\label{fig:basic}}
\end{figure}

The noninteracting part of Eq.~\ref{eq_H} can be solved for the energy eigenvalues. For physically realistic negative tunneling energies (i.e., $t>0$) there are three bands, as shown in Fig.~\ref{fig:basic}(a).  The highest band is flat (dashed line).  At a density of 2/3 we fill the lowest two bands (solid lines).  Here the non-interacting Fermi surface touches the empty flat band at a QBCP [red arrow in Fig.~\ref{fig:basic}(a)]. We will see that the dipolar interaction opens a gap at the QBCP.

To construct the phase diagram of Eq.~\ref{eq_H} with $V_1>0$ we use two complementary methods:  MFT and ED.  ED includes all quantum fluctuations but applies only to small system sizes. Specifically, we use the Krylov-Schur algorithm \cite{Stewart2001} which allows us to handle degenerate eigenvalues.  This method is essentially exact because it is unbiased and gives the same results as other unbiased methods on small lattices.  With ED we work on a finite system size, 27 sites ($3\times3$ unit cells) and $N=18$ fermions, with periodic boundaries to obtain the lowest energy states.

The MFT we use, in contrast, applies to either finite or infinite system sizes.  It excludes quantum fluctuations due to our choice for decoupling of the interactions.  The following Hartree-Fock decoupling turns out to be surprisingly accurate in comparison to ED:
\begin{eqnarray}
n_{\bm{r}}n_{\bm{r'}} & \rightarrow  &
\bar{n}_{\bm{r}} n_{\bm{r'}}
+n_{\bm{r}}\bar{n}_{\bm{r'}} 
- \bar{n}_{\bm{r}}\bar{n}_{\bm{r'}}\nonumber\\
& & -\psi_{{\bm{r'}},{\bm{r}}}c_{\bm{r}}^{\dagger}c_{\bm{r'}}^{\vphantom{\dagger}}
-\psi_{{\bm{r}},{\bm{r'}}}
c_{\bm{r'}}^{\dagger}c_{\bm{r}}^{\vphantom{\dagger}}
+\vert \psi_{{\bm{r'}},{\bm{r}}}\vert^2
\end{eqnarray}
where the fields that capture circulating currents and density oscillations are $\psi_{{\bm{r'}},{\bm{r}}}\equiv \left\langle c_{\bm{r'}}^{\dagger}c_{\bm{r}}^{\vphantom{\dagger}}\right\rangle$ and $\bar{n}_{\bm{r}}\equiv\left\langle n_{\bm{r}}\right\rangle$, respectively.

The decoupling of the interaction in Eq.~\ref{eq_H} leads to a quadratic Hamiltonian which we solve self-consistently. We then obtain the finite temperature phase diagram from the free energy at temperature $T$:
\begin{eqnarray}
F  &=&  -T\sum_{\bm{k},m}\log\left[1+e^{\left(\mu-E_{\bm{k}m}\right)/T}\right]
  +\mu N\nonumber\\
 & &+\frac{V_1}{2}\sum_{\bm{r}\neq\bm{r'}}
 \frac{\vert\psi_{{\bm{r'}},{\bm{r}}}\vert^2
 -\bar{n}_{\bm{r}}\bar{n}_{\bm{r'}}}{|\bm{r}-\bm{r'}|^3},
\end{eqnarray}
where $E_{\bm{k}m}$ are the single-particle eigenvalues of the mean-field quadratic terms in the decoupled Hamiltonian in the $m^{th}$ band at wavevector ${\bm{k}}$. The chemical potential, $\mu$, is determined by requiring $\partial F/\partial\mu=N$.  We find the lowest free energy by starting with random initial guesses for the fields $\psi_{{\bm{r'}},{\bm{r}}}$ and $\bar{n}_{\bm{r}}$ and self-consistently iterating. 
With the truncation of $\left|{\bm r}-{\bm r'}\right|<5$, each site interacts with 62 neighbors, and hence there are 3 independent real values of $\bar{n}_{\bm r}$ and 93 independent complex values of $\psi_{{\bm{r'}},{\bm{r}}}$ which have to be solved self-consistently.
From these solutions we obtain the finite temperature phase diagram as well as other thermodynamic functions, such as the entropy per particle, $s= -N^{-1}\partial F/\partial T$.

\section{Results}\label{sec:results}
Our analysis of Eq.~\ref{eq_H} finds two competing orders in MFT.  At large interaction strengths we expect the long-range dipolar interaction to establish a CDW.  We characterize a CDW by long-range oscillations in the density such that  
$
\delta n=\max_{\bm{r}}(\bar{ n}_{\bm{r}}) -\min_{\bm{r}}(\bar{n}_{\bm{r}})
$
is non-zero.  Fig.~\ref{fig:basic}(b) shows the stripe-CDW pattern we find in MFT. 

We also expect that the frustrated lattice will penalize CDWs and allow competition from uniform phases ($\delta n=0$).  We find that interactions generate a uniform TMIP that spontaneously breaks time-reversal symmetry to generate loop currents:
\begin{eqnarray}
I\equiv 2\max_{\bm{r},\bm{r'}}\vert{\rm Im} \psi_{{\bm{r'}}{\bm{r}}}\vert.
\end{eqnarray}
Here $I>0$ indicates a phase with non-zero bond current.  Fig.~\ref{fig:basic}(c) shows the current pattern we find in the TMIP.  Here we see that the outside edge maintains a chiral current. The direction of edge flow is spontaneously chosen.  The bulk gap and edge chiral currents establish a quantized Hall effect in the absence of an applied field, i.e., a QAH effect. 

\begin{figure}
\begin{centering}
\includegraphics[width=\columnwidth]{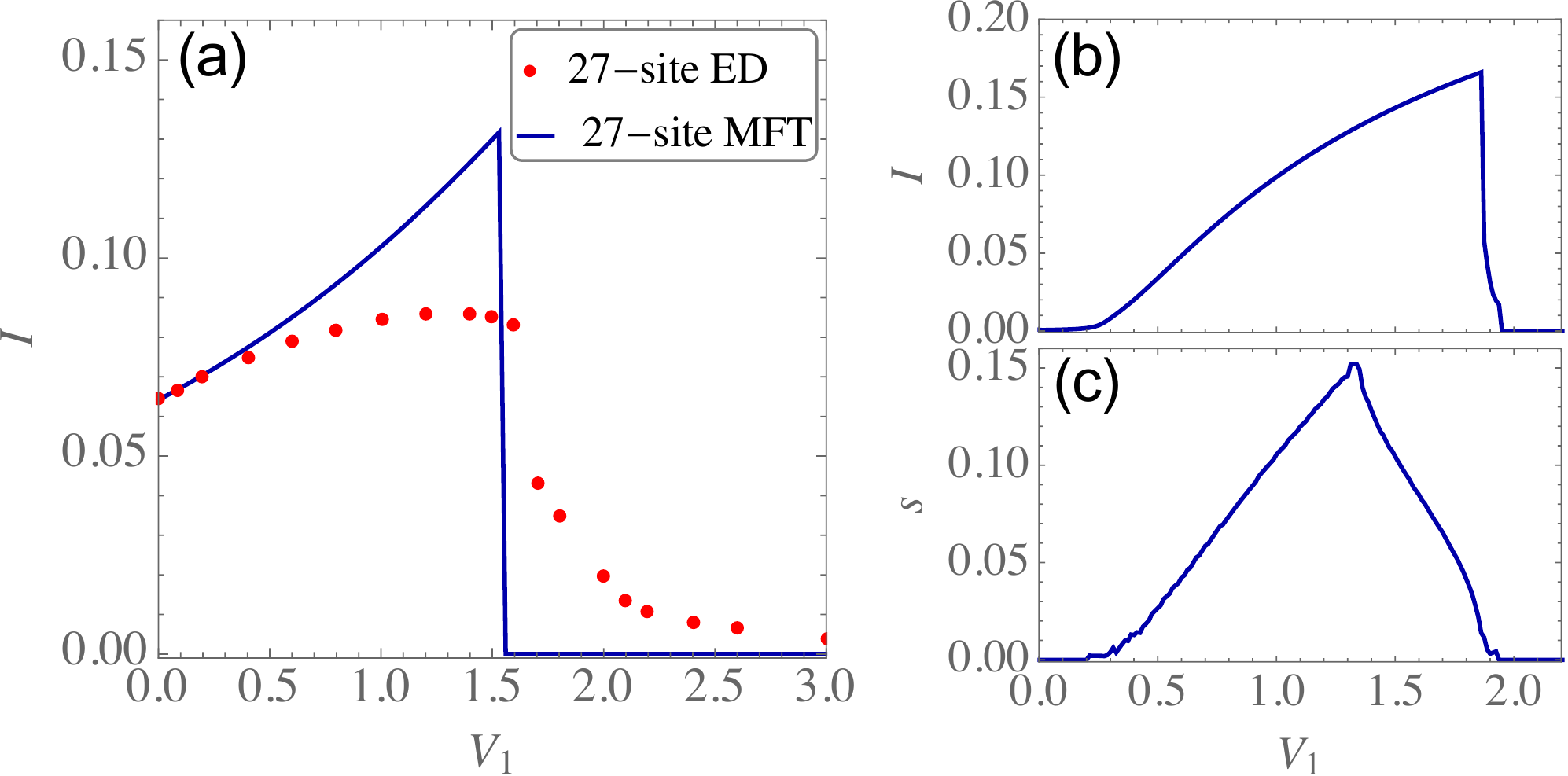}
\par\end{centering}
\caption{(a) Current plotted against the nearest neighbor interaction strength at zero temperature on a 27-site kagome lattice, 
obtained from exact diagonalization (dots) and mean-field theory (line) applied to Eq.~\ref{eq_H}.  Non-zero current implies a topological Mott insulator with a quantum anomalous Hall effect. Here finite-size effects (see Appendix~\ref{app:current}) lead to a non-zero current at $V_1=0$.  Both methods capture the transition from the topological Mott insulator  (small $V_1$) to a charge density wave (large $V_1$). 
(b) Current plotted against the interaction strength at zero temperature on an infinite system using mean-field theory on Eq.~\ref{eq_H} showing the absence of current at $V_1=0$ and the same transition as in (a). 
(c) The same as (b) but plotting the maximum attainable entropy per particle.\label{fig:IandS}}
\end{figure}

Quantum fluctuations excluded in MFT may favor the CDW over the TMIP.  To test the stability of the TMIP against quantum fluctuations we employ ED on finite system sizes and compare with MFT.  We find that ED and MFT produce nearly the same low energy manifold with precisely the same configurations of order parameters, as shown in Fig.~\ref{fig:basic}(b) and (c).  We also compare the transition between TMIP and CDW found from both methods.  Fig.~\ref{fig:IandS}(a) plots the current versus nearest neighbor interaction strength for both MFT and ED for 27 sites.  Here we see that MFT and ED are exactly the same for the non-interacting case, as expected.  The non-zero $I$ at $V_1=0$ arises because of a finite size effect (see Appendix~\ref{app:current}).  Fig.~\ref{fig:IandS}(a) also shows that the TMIP transitions to a CDW at large interactions in both methods.  The agreement shows the remarkable accuracy of MFT in predicting the structure and magnitude of the order parameters, low energy Hilbert space, and location of phase transitions. 

Fig.~\ref{fig:IandS}(b) plots the same as (a) but in the thermodynamic limit using MFT.  Here we see that the chiral current is zero in the absence of interactions in the thermodynamic limit.  Interaction strengths on the order of the tunneling trigger spontaneous chiral currents in the TMIP. 

We now turn to the thermodynamics of the TMIP.  In the absence of heat and particle number reservoirs, the temperature of atomic and molecular gases placed in optical lattices is set by the entropy.  Fig.~\ref{fig:IandS}(c) plots the entropy needed to cool into the TMIP obtained by MFT.  Here we see that the TMIP is most stable near $V_1\approx1.3$, where the gap is the largest, establishing the highest critical entropy per particle to be $s_{c}\approx0.15$.  

\begin{figure}
\begin{centering}
\includegraphics[width=1\columnwidth]{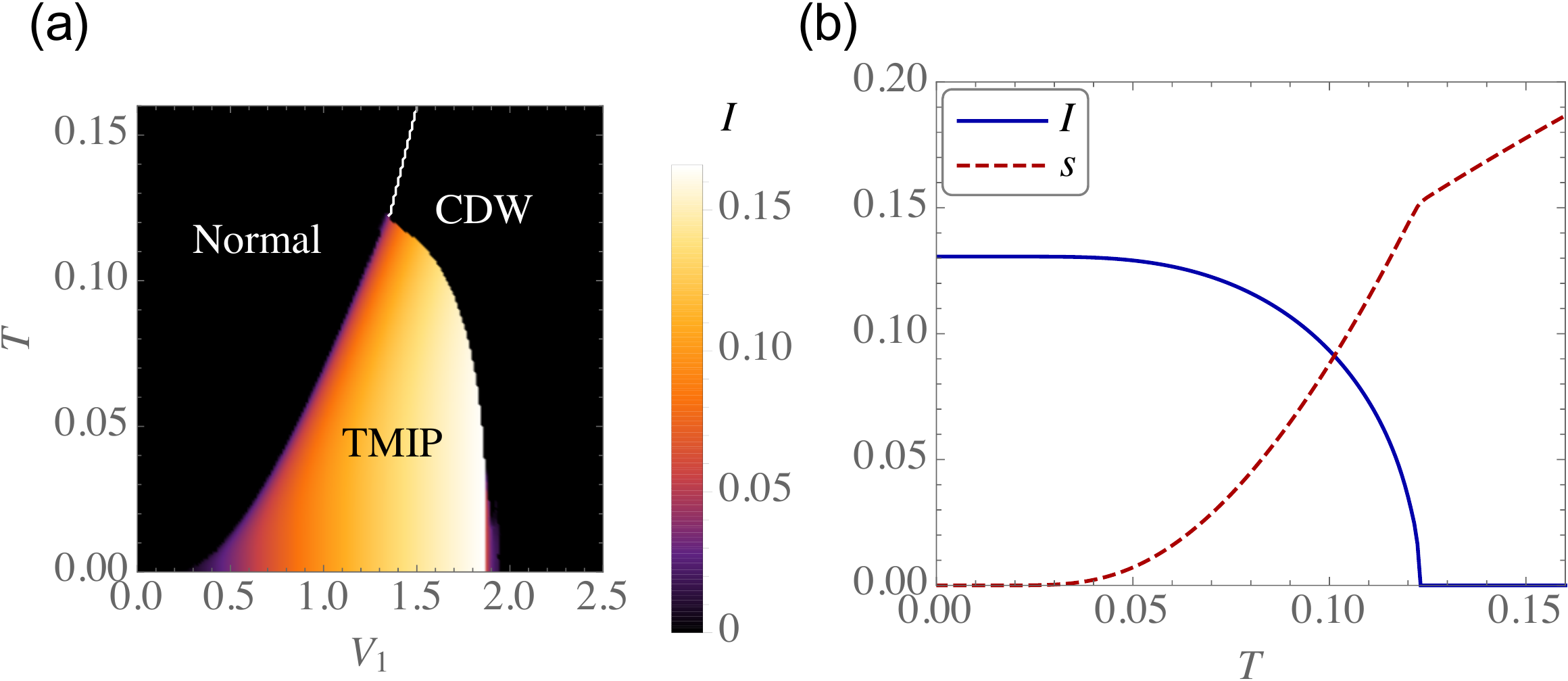}
\par\end{centering}
\caption{(a) The mean-field phase diagram of Eq.~\ref{eq_H} obtained by plotting the magnitude of the current against temperature and interaction strength.  
The white line uses the density difference between sites, $\delta n$, to plot the boundary between the charge density wave ($\delta n>0$) and the Normal phase (an absence of order with $\delta n=0$).
(b) Current and entropy per particle plotted against temperature
where the quantum anomalous Hall effect is strongest, $V_{1}=1.33$. The kink in the entropy shows a first order phase transition from the topological Mott insulator to a charge density wave.\label{fig:T}}
\end{figure}

Thermal fluctuations drive transitions out of the TMIP. Fig.~\ref{fig:T}(a) plots the full finite temperature phase diagram of Eq.~\ref{eq_H} obtained from MFT.  With increasing temperature we see two types of thermal phase transitions.  The TMIP either undergoes a second-order
phase transition to the Normal phase (for $V_{1}\lesssim1.33$) or
a first-order transition to the CDW phase (for $1.33\lesssim V_{1}\lesssim1.9$).  Here the Normal phase is define by an absence of order. 
The highest critical temperature of the TMIP is $T_{c}\approx0.12$, indicating that the bi-critical
point (in-between the Normal, TMIP and CDW phases) is at $(V_{1c},T_{c})\approx(1.33,0.12)$.

The phase diagram shows that lower entropies will improve stability and observability of the QAH effect.  The current $I$ is a key observable that becomes enhanced at low temperatures.  Fig.~\ref{fig:T}(b) plots both the current and entropy as a function of temperature for the interaction strength where the TMIP is the strongest, $V_1=1.33$.  Here we see that the current is essentially zero for entropies per particle above $\approx0.15$.  But at lower entropies the current becomes observable thus signaling the TMIP.  

\section{Discussion}\label{sec:discussion}

We propose that the TMIP in the kagome optical lattice arises from just dipolar interactions. We argue that even in the absence of applied fields it should display a QAH effect as a result of spontaneous time reversal symmetry breaking.  Prospects for realizing the TMIP therefore offer a key advantage over other proposals to realize topological phases based on applied synthetic fields because the TMIP will not have as much heating or losses due to these additional fields.  Furthermore, the crucial ingredients to realizing TMIP have already been experimentally implemented: a dipolar interaction and a deep kagome optical lattice.  Yet there are other potential experimental challenges.  The small gap of the TMIP leaves it somewhat sensitive to trapping and heating. Using MFT we can estimate the impact of these realistic effects.

We must first estimate the gap in a realistic setting to establish the  overall stability of the TMIP. In the grand canonical ensemble the chemical potential may then vary within the gap while preserving the QAH effect.  The gap is in turn set by the ratio of the tunneling and interaction strength.  

To estimate the gap we first model the optical lattice potential to accurately obtain the Wannier functions and the tunneling.  For a kagome lattice generated by three pairs of long- and short-wavelength lasers\cite{Jo2012}
with lattice length $355{\rm nm}$ and depth (defined in Appendix~\ref{app:kagomeFormula})
$7.8E_{R}$, the nearest-neighbor tunneling can be estimated using a Gaussian
approximation for the Wannier functions.  We find $t\sim0.009E_{R}$.  We have verified the Gaussian approximation at these lattice depths by computing
the band dispersion through a plane-wave expansion and comparing with
the bandwidth of a tight-binding model on a kagome lattice with only nearest-neighbor tunneling.

To estimate the nearest-neighbor interaction we consider an example atom with a strong dipolar moment which has already been cooled to quantum degeneracy: $^{161}{\rm Dy}$ \cite{Lu2012}.  At the lattice length of $355{\rm nm}$, we find $V_{1}\approx0.012E_{R}$, assuming perfectly localized Wannier functions (the correction due to the finite spreads of Wannier functions is small, see Appendix~\ref{app:dipolar}).  Therefore, the lattice depth of $7.8E_{R}$ gives  $V_{1}/t\sim1.3$, which is the optimal point for the TMIP in the mean-field phase diagram since the TMIP has the highest gap here.  Using MFT we find a gap of $\Delta\approx0.46t$ at these lattice depths. 
 
The gap determines the robustness against perturbations such as confinement.  Assuming parabolic confinement, of strength $M(\omega r)^{2}/2$, where $\omega$ is the trapping frequency and $M$ is the mass of $^{161}{\rm Dy}$,  we can estimate the spatial extent of the TMIP  by assuming that the TMIP survives until the trap strength equals the gap, i.e., $
M(\omega r)^{2}/2=\Delta
$. 
A gap $\Delta\approx0.46t$ 
with trap strength $\omega\sim 2\pi \times 10$Hz, leaves a TMIP about 20 sites in diameter.

The size of the gap also sets the thermal stability of the TMIP.  Conventional evaporative cooling in a harmonic trap can cool to entropies per particle as low as $\approx 0.25$ ($\approx 0.75$) for bosons (fermions) or possibly lower \cite{McKay2011,onofrio2017}.  Whereas more recent results with atomic gas microscopes cooling into the antiferromagnetic phase of the two-dimensional Hubbard model have reached entropies per particle lower than $0.75$ for fermions \cite{Mazurenko2017}.  The entropies per particle  required to reach the TMIP  ($\approx 0.15$) with dipolar fermions are therefore potentially within reach of current experiments with atomic gas microscopes. Nonetheless, careful preparation of a reservoir \cite{Mazurenko2017} will be needed to reach these low entropies. 

Once prepared, the topological phase can be detected by its chiral edge currents. A number of proposals have been put forth for the direct detection of topological properties
\cite{Zhao2011,Goldman2012,wang2013,Goldman2013,Spielman2013,Goldman2016a,Price2016,Tran2017} with several successful experimental implementations \cite{Leblanc2012,Atala2014,Mancini2015,Stuhl2015,Nakajima2016,Lohse2016,Tai2017}.  For example, recent experiments with atomic gas microscopes have been able to directly observe chiral edge states in a Hofstadter band thus offering a direct route to detecting the QAH effect derived from the TMIP\cite{Tai2017}.  Once established, a TMIP would set the stage for possible detection of anyons in fractional TMIPs \cite{Zhu2016a}.

VWS, MC and HH acknowledge support from AFOSR (Grant No. FA9550-15-1-0445), ARO (Grant No. W911NF-16-1-0182), and Advanced Research Computing at Virginia Tech for providing computational resources. ST acknowledges support from ARO (Grant No. W911NF-16-1-0182).

\appendix

\section{Kagome Optical lattice potential}\label{app:kagomeFormula}

We use the kagome optical lattice potential implemented in Ref.~\onlinecite{Jo2012}.  The potential experienced by atoms is given by:
\begin{equation}
V_L({\bf r})=V_0\sum_{n=1}^3 \left[\sin^2(\frac{2\pi}{\sqrt{3}a}{\bf r}\cdot{\bf d}_n)
           - \sin^2(\frac{\pi}{\sqrt{3}a}{\bf r}\cdot{\bf d}_n)\right]
\label{eq_Vpotential}
\end{equation}
where $a$ is the distance between adjacent sites,
${\bf d}_n=\cos\frac{2n\pi}{3}\hat{x}+\sin\frac{2n\pi}{3}\hat{y}$,
and $V_0$ is the lattice depth.

\section{Dipolar interaction}\label{app:dipolar}

The model Hamiltonian, given by Eq.~(1),
implies that the interaction strength decays with respect to distance $r$ as $r^{-3}$.
This is not strictly true for a realistic system at short range since the Wannier functions have a finite spread. 
For the lattice depth of $2.3 E_R$, as discussed in the main text,
we use a Gaussian approximation to find the Wannier functions and thereby compute
the real interaction strengths, $V_n$, for the $n^{\text{th}}$-nearest neighbor.
Their ratios to those estimated from simplistic $1/r^3$ are given in  Table~\ref{table:Vn}.

We have checked that the results presented in the main text are consistent with these revised values for the interaction.  For example, we find that perturbing the interaction to the values in Table~\ref{table:Vn} leads to less than one percent shift in the critical temperature of the TMIP in the mean-field approach.  We therefore conclude that the TMIP gap leaves it robust enough to use the approximate interaction discussed in Eq.~(1). 

\begin{table}
\begin{centering}
\begin{tabular}{|c|c|}
\hline 
$n$ nearest neighbor & Real $V_{n}$ / Estimated $V$\tabularnewline
\hline 
\hline 
1 & 1.200\tabularnewline
\hline 
2 & 1.045\tabularnewline
\hline 
3 & 1.030\tabularnewline
\hline 
\end{tabular}
\par\end{centering}
\caption{Table of ratios of real $V_{n}$ and estimated $V_{n}$, for a number
of $n$}
\label{table:Vn}
\end{table}
 
 \vspace{1cm}
\section{Nonzero Current in the Non-interacting limit In a Finite- size System}\label{app:current}

Fig.~3 of the main text shows a non-zero current in the non-interacting limit in a finite-size system. This can be understood as a finite size effect within the non-interacting band structure. With reference to Fig.~2(a) of the main text, in the absence of interactions the ground state can be considered as filling the single-particle levels up to the Fermi level. At the filling ratio of $2/3$, the last particle has the freedom to occupy either the $\Gamma$-point of the second band [red arrow of Fig.~2(a)] or any state in the topmost flat band, all with the same energy. Infinitesimal interaction, however, favors a finite-momentum state of the topmost state to be occupied, and thus we find finite current in the limit $V_1\rightarrow0$.

The current contributed from the last particle is appreciable in the finite-size systems only. Since the current is computed as $\left<I\right>=N^{-1}\sum_i\left<i\right|\hat{I}\left|i\right>$ (where $\left|i\right>$ are the occupied states), in the thermodynamic limit the current at $V_1=0$, contributed from the last particle alone, is suppressed by $N^{-1}$.

\bibliography{library}

\end{document}